\documentclass[superscriptaddress,onecolumn,amssymb,amsmath,nobibnotes,aps,prd,showkeys,nofootinbib]{revtex4}

\makeatletter

\renewcommand*{\p@subsection}{}

\renewcommand*{\p@subsubsection}{}
\makeatother

\usepackage[Export]{adjustbox}
\usepackage{amsmath, amssymb, amsfonts, amsthm}
\usepackage{epsfig}
\usepackage{graphicx}  
\usepackage[utf8]{inputenc}
\usepackage{color}     
\usepackage{lscape}
\usepackage{color}

\begin{document}

\title{Modified Emergent Dark Energy and its Astronomical Constraints}

\author{H. B. Benaoum} 
\email{hbenaoum@sharjah.ac.ae}

\affiliation{Department of Applied Physics and Astronomy,
University of Sharjah, United Arab Emirates}

\author{Weiqiang Yang}
\email{d11102004@163.com}
\affiliation{Department of Physics, Liaoning Normal University, Dalian, 116029, People's Republic of China}

\author{Supriya Pan}
\email{supriya.maths@presiuniv.ac.in}
\affiliation{Department of Mathematics, Presidency University, 86/1 College Street, Kolkata 700073, India}

\author{Eleonora Di Valentino}
\email{eleonora.divalentino@manchester.ac.uk}
\affiliation{Jodrell Bank Center for Astrophysics, School of Physics and Astronomy, 
University of Manchester, Oxford Road, Manchester, M13 9PL, UK.}

\begin{abstract}
We introduce a modified form of the Phenomenologically Emergent Dark Energy (PEDE) model by showing a very elegant approach. The model is named as Modified Emergent Dark Energy (MEDE) to distinguish from PEDE model and  it includes $\Lambda$CDM, PEDE model, Chevallier-Polarski-Linder model and other cosmological models of interest. We show that the present article offers a very fantastic route to construct other PEDE models in a simple but very elegant way. The model has seven free parameters where the six parameters are the same as in $\Lambda$CDM or PEDE model and the remaining one parameter `$\alpha$' quantifies the generalization. 
The present model predicts that dark energy equation of state at present assumes,
 $w_{\rm DE}\; (z=0) = -1 - \frac{\alpha}{3 \ln (10)}$ and in the far future (i.e., for $z \longrightarrow -1$), it will evolve asymptotically to $w_{\rm DE} \longrightarrow -1$. We perform a very robust observational analysis of the model using several observational datasets including cosmic microwave background radiation, baryon acoustic oscillations distance measurements, a local value of the Hubble constant and the pantheon sample of Supernovae Type Ia. We find that the $H_0$ tension here is alleviated but not solved, while the only free parameter, $\alpha$ could recover the $\Lambda$CDM or PEDE for different datasets. In summary, the present article describes a novel route to construct other modified versions of the PEDE model in a very simplified manner. 
\end{abstract}
\maketitle
\section{Introduction}

Over the years, cosmologists are trying to understand the mysterious accelerating phase of the Universe in the recent time with the help of various astronomical data. Since the very end of nineties, the scientific community has witnessed the arrival of many new theories to explain this accelerating phase, however, none of them are found to be super excellent for explaining this late accelerating phase. Therefore, we are being introduced with new cosmological models 
aiming to improve the existing models and theories. If we look at the existing literature, one can clearly see that there are already different theoretical proposals. Two well known theories in this context are the theory of dark energy (within the context of General Relativity) \cite{Copeland:2006wr,Huterer:2017buf} and the Modified gravitational theories \cite{Sotiriou:2008rp,DeFelice:2010aj,Capozziello:2011et,Cai:2015emx,Nojiri:2017ncd}. 
In dark energy theory, we modify the matter content of the Universe in the framework of the Einstein's General Relativity (GR) by introducing a hypothetical fluid with large negative pressure so that the accelerating expansion of the Universe is described. On the other hand, modified gravity theories are in search of some alternative gravitational theory beyond GR where the geometry of the space-time could lead to this observed acceleration. Both the approaches achieved significant attention in our scientific community, and as a consequence, the present literature has been extremely heavy with a number of dark energy and modified gravitational models. Despite of many interesting models from both ends, the search for a perfect cosmological model describing the observational evidences quite accurately is still going on. 
In the present article we consider the first approach, that means, we focus on the dark energy fluid which according to a series of observational evidences occupies nearly 68\% of the total energy budget of the Universe. So far we are aware of the dark energy physics, its nature, dynamics and origin are totally unknown. Therefore, theoretically one can try to understand the accelerating expansion by introducing different versions of the dark energy models. The present work is motivated in a similar fashion but with a novel idea which aims to generalize a recently introduced cosmological model which very soon attracted the scientific community for its potentiality to address one of the striking problems in modern cosmology, the Hubble constant $H_0$ tension, without using any additional degrees of freedom. The Hubble constant tension is one of the very serious issues in modern cosmology which dictates that $\Lambda$CDM is probably not the best solution for the present observed universe and perhaps we have to widen our mind to accept alternative cosmological models beyond the mathematically simplest cosmological model $\Lambda$CDM. Let us digress a bit towards the issue known as Hubble constant tension.  
The $H_0$ tension is the disagreement existing between the model dependent estimates of the Hubble constant using CMB or BAO data~\cite{Aghanim:2018eyx,Aiola:2020azj,Alam:2020sor} and the direct local measurements~\cite{Riess:2019cxk,Reid:2019tiq,Wong:2019kwg,Pesce:2020xfe}. This discrepancy found in both the estimations are quite significant to look at very seriously. 
There are plenty of tentative solutions in the literature, mostly based on the models of Dark Energy~\cite{Yang:2018qmz,Yang:2018prh,DiValentino:2019dzu,Vagnozzi:2019ezj,DiValentino:2020naf,Keeley:2019esp,Joudaki:2016kym,Poulin:2018cxd,Karwal:2016vyq,Sakstein:2019fmf,Niedermann:2019olb,Ye:2020btb,Agrawal:2019lmo,Lin:2019qug,Berghaus:2019cls,Smith:2019ihp,Lucca:2020fgp,Li:2019yem,Pan:2019hac,Rezaei:2020mrj,Liu:2020vgn,Li:2020ybr,Yang:2020tax}, or Interacting Dark Energy
~\cite{Kumar:2016zpg,DiValentino:2017iww,Kumar:2017dnp,vandeBruck:2017idm,Yang:2018euj,Yang:2018uae,Yang:2019uzo,Kumar:2019wfs,Pan:2019jqh,Martinelli:2019dau,DiValentino:2019ffd,DiValentino:2019jae,Yang:2019uog,Benevento:2020fev,Gomez-Valent:2020mqn,Lucca:2020zjb,Agrawal:2019dlm,Anchordoqui:2019amx,Anchordoqui:2020sqo,Pan:2020zza,Pan:2020bur}, or Modified Gravity~\cite{Nunes:2018xbm,Raveri:2019mxg,Yan:2019gbw,Frusciante:2019puu,DAgostino:2020dhv,Wang:2020zfv} or many others possibilities~\cite{Hart:2017ndk, Chiang:2018xpn, Hart:2019dxi, Yang:2019jwn, Jedamzik:2020krr, Sekiguchi:2020teg,Bose:2020cjb,DiBari:2013dna,Berezhiani:2015yta,Anchordoqui:2015lqa,Vattis:2019efj,Desai:2019pvs,Alcaniz:2019kah,Chudaykin:2016yfk,Yang:2019qza,Chudaykin:2017ptd,Kreisch:2019yzn,Blinov:2019gcj,DiValentino:2017oaw,Yang:2020zuk,Nunes:2020uex}, however, based on the existing literature cited above, none of them are fully satisfactorily solving this issue when more datasets are taken into account~\cite{Knox:2019rjx}. Therefore, alternative cosmological models are on high demand to understand the physics of the $H_0$ tension.

So, as already mentioned, there is no reason to restrict ourselves to some well known cosmological models. Since it has been already found  over the years that the simplest cosmological model, the $\Lambda$CDM, has been diagnosed with some severe problems. Starting from the cosmological constant problem to the recently detected $H_0$ tension, $\Lambda$CDM cosmology has been kept under the microscope over and over. Scientific mind always demands fairness, so the new cosmological models are equally welcome. However, since the observational science has developed a lot, so, a proposed theory should always be tested with them in order to examine whether the proposed theory should be included in the literature or it must be thrown right away. The present article thus starts with an unbiased mind where we recall a new cosmological model in the name of  Phenomenologically Emergent Dark Energy (PEDE), proposed recently in \cite{Li:2019yem} in a spatially flat Friedmann-Lema\^{i}tre-Robertson-Walker (FLRW) Universe. The model is symmetrical around the present time and within this model scenario, dark energy emerges only in the late time but it has no presence in the past. 
The most interesting feature of the model is that, it has no extra degrees of freedom similar to the six parameters based $\Lambda$CDM model. One of the intriguing properties of the model is that, it can alleviate the $H_0$ tension within 68\% CL \cite{Pan:2019hac}. The model seems to be very effective because in an extended cosmological scenarios including the presence of massive neutrinos, this model has been found to be quite able to alleviate the $H_0$ tension \cite{Yang:2020myd}. Naturally, the potentiality of the PEDE model has been recognized shortly \cite{Li:2019yem,Pan:2019hac,Yang:2020myd}.

Nevertheless, the curious mind cannot stop here and it always tries to generalize any theory and model. According to the past historical records, attempts to generalize any theory have been taken earlier by many cosmologists. The present work follows the earlier strategy where we generalize the PEDE model in the name of {\it Modified Emergent Dark Energy} (MEDE). This model can recover the PEDE model as a special case. We investigate the behaviour of the model in terms of various cosmological variables. 
We then constrain the model using a series of recently available cosmological datasets including Cosmic Microwave Background from Planck 2018 legacy release, Baryon Acoustic Oscillations distance measurements, a local measurement of the Hubble constant from the Hubble Space Telescope, Pantheon sample from the Supernovae Type Ia.

The article is organized in the following way. In section \ref{sec-mede} we present the MEDE model and its cosmological features. After that, in section \ref{sec-observational data}, we describe the observational data and methodology to constrain the MEDE model. Section \ref{sec-results} describes the observational constraints on the model and finally, in section \ref{sec-conclu}, we close the present work.

\section{Modified Emergent Dark Energy Model}
\label{sec-mede}

According to the past and present observational surveys, in the large scale, our Universe is almost homogeneous and isotropic.  This homogeneous and isotropic characterization of the Universe is well described by the spatially 
flat Friedmann-Lemaitre-Robertson-Walker (FLRW) line element:
$ds^2  = -dt^2 + a^2 (t) (dx^2 +dy^2+dz^2)$, written in terms of the comoving coordinates $(t, x, y, z)$ where $a (t)$ is the dimensionless expansion scale factor of the Universe. We assume that the gravitational sector of the Universe follows the Einstein General Relativity, where the matter distribution of the Universe is minimally coupled to the gravitational sector and none of the fluids in the matter sector is interacting with other fluid, that means effectively we are considering a purely non-interacting cosmological scenario. Within such framework,  one can write down the 
Hubble function $H (z) = \dot{a}/a$ (the dot is taken with respect to the cosmic time) in the spatially flat Universe as function of the redshift $z$, as follows 
\begin{eqnarray}
H (z)^2  & = & H_0^2 \left[ \Omega_{r0} (1+z)^4 + \Omega_{m0} (1+z)^3 + \tilde{\Omega}_{DE} (z) \right],
\end{eqnarray} 
where $\Omega_{r0}$ and $\Omega_{m0}$ are respectively the density parameters for the radiation and the matter sector at present time; $\tilde{\Omega}_{\rm DE} (z)\; (= \rho_{\rm DE}/\rho_{\rm crit,0} )$ is the dark energy density normalized with respect to the critical density and solving the usual conservation equation for the dark energy having no-interaction with others, one could express $\tilde{\Omega}_{\rm DE} (z)$ as  
\begin{eqnarray}
\tilde{\Omega}_{\rm DE} (z) & = & \Omega_{\rm DE,0} ~\exp \left(3 \int_0^z \frac{1+w_{\rm DE} (z')}{1+z'} d z' \right),
\end{eqnarray}
where $\Omega_{\rm DE,0} = 1 - \Omega_{r0} - \Omega_{m0}$ is the current value of the dark energy density parameter and $w_{\rm DE} (z) = p_{DE} (z)/\rho_{\rm DE} (z)$ is the equation of state (EoS) of the dark energy fluid. Following the earlier works \cite{Li:2019yem,Pan:2019hac}, here we are interested to parametrize $\tilde{\Omega}_{\rm DE} (z)$ but in a generalized way. 

In order to generalize the earlier works \cite{Li:2019yem,Pan:2019hac}, we express the normalized dark energy density $\tilde{\Omega}_{\rm DE} (z)$ as:
\begin{eqnarray}\label{MEDE0}
\tilde{\Omega}_{\rm DE} (z) & = & \Omega_{\rm DE,0} ~G (z), 
\end{eqnarray}
where $G (z)$ is a generic function equal to $1$ at present time (i.e. $z=0$), which we parametrize having the following functional form: 
\begin{eqnarray}\label{MEDE}
G (z)  & = & \frac{2}{1 + g (z)^{\frac{2}{\ln (10)}}} 
= 1 -\tanh \left[\log_{10} \left( g(z)\right) \right], 
\end{eqnarray}
where $g (z): \mathbf{R} \longrightarrow \mathbf{R}$ is a continuous real function satisfying the condition $g (z=0) = 1$, which can be determined later. One can clearly see that the model (\ref{MEDE}) is the generalized version of the PEDE model \cite{Li:2019yem,Pan:2019hac,Yang:2020myd}, because with the choice of $g (z) = 1+z$, in eqn. (\ref{MEDE}), PEDE model is recovered as a special case. Additionally, the $\Lambda$CDM model corresponds to $g (z) = 1$ and the Chevallier-Polarski-Linder parametrization \cite{Chevallier:2000qy,Linder:2002et} corresponds to $g(z)^{2/\ln(10)}= -1 + 2 (1+z)^{-3 (1+w_0+w_a)} ~\exp \left[(3 z w_a)/(1+z) \right]$. 
Due to the flexibility with the introduction of $g (z)$ in eqn. (\ref{MEDE}) one can choose any specific functional form for $g(z)$ and introduce a class of cosmological models in this direction. The model  (\ref{MEDE0}) for the choice of $G (z)$ in eqn. (\ref{MEDE}) is the Modified Emergent Dark Energy  (MEDE) model in this work. 
Now, for the normalized dark energy density,  $\tilde{\Omega}_{\rm DE} (z)$, using the conservation equation $\dot{\rho}_{\rm DE} + 3 H (1+w_{\rm DE}) \rho_{\rm DE} = 0$, one can determine the dark energy equation-of-state as 

\begin{eqnarray}
w_{\rm DE} (z) & = & -1 + \frac{1}{3} (1+z) ~\frac{d \ln \tilde{\Omega}_{\rm DE} (z)}{dz}, 
\end{eqnarray}
which for the modified emergent dark energy (MEDE) model takes the form:
\begin{eqnarray}
w_{\rm DE} (z) & = & -1 - \frac{1}{3 \ln (10)}  \frac{(1+z)}{g(z)}~\frac{d g (z)}{d z} ~~ \left(1+ \tanh \left[\log_{10} \left( g(z)\right) \right] \right), \nonumber \\
& = & - 1 - \frac{1}{3 \ln (10)} \frac{(1+z)}{g(z)}~\frac{d g (z)}{d z} \left(2 - G (z) \right),
\end{eqnarray}
Now, at this point, to proceed with the dynamics of such model, it is necessary to choose a specific form of the function $g(z)$.
To determine the function $g (z)$, we impose a simple condition on the EoS of the dark energy which states that $w_{\rm DE} (z)$ as a function depending explicitly on $G (z)$ only. This means that:
\begin{eqnarray}\label{condition}
 \frac{(1+z)}{g(z)}~\frac{d g (z)}{d z} & = & {\rm constant} = \alpha
\end{eqnarray}
where $\alpha$ is an integration constant which could be either positive or negative. Below we shall discuss how the sign of $\alpha$ becomes important for characterizing the emergent nature of the present dark energy. 
Now, the above equation (\ref{condition}) determines uniquely the function $g (z)$, which in this case takes the form:
\begin{eqnarray} \label{alpha}
g (z) & = & (1 + z)^{\alpha}~. 
\end{eqnarray}
Note that we recover the $\Lambda$CDM model for $\alpha=0$, and the PEDE model for $\alpha=1$. With the above function in eqn. (\ref{alpha}), 
the generic function $G (z)$ for the MEDE model now becomes,
\begin{eqnarray}
G (z) = 1 -\tanh \left[\alpha \log_{10} \left(1+z\right) \right], 
\end{eqnarray}
and therefore, one can recast the Hubble equation as,
\begin{eqnarray}
 E (z) & = & \frac{H (z)}{H_0} =\left[ \Omega_{r0} (1+z)^4 + \Omega_{m0} (1+z)^3 + 
\Omega_{\rm DE,0} \left( 1 -\tanh \left[\alpha \log_{10} \left(1+z\right) \right] \right) \right]^{1/2}. 
\end{eqnarray} 
The EoS of dark energy for the above choice of $g(z)$ can also be written as:
\begin{eqnarray}
w_{\rm DE} (z) = -1 - \frac{\alpha}{3 \ln (10)} ~\Bigl(1+ \tanh \bigl[\alpha \log_{10} \left(1+z \right) \bigr] \Bigr).  
\end{eqnarray}

We should note that one may consider a more general condition than the one, i.e., eqn. (\ref{condition}), we used to determine the function $g (z)$ and consequently obtain other emergent like dark energy, for instance \cite{Li:2020ybr}. {\it This is the potentiality of this model which dictates that the condition on $g (z)$ and its derivative is the building block of many new models. }
One can also see that the  in the far future (i.e., $z \longrightarrow -1$), the dark energy equation of state will evolve asymptotically to $w_{\rm DE} \longrightarrow -1$. At present time, the model predicts that: 
\begin{eqnarray} \label{DEeos}
w_{\rm DE} (z=0) & = & -1 - \frac{\alpha}{3 \ln (10)},
\end{eqnarray}
and this could be phantom or quintessence depending on the sign of $\alpha$. For $\alpha > 0$, $w_{\rm DE} < -1$ while $w_{\rm DE} > -1$ for $\alpha < 0$. 
Now, one could be more generalized on the dark energy EoS at present time through the generalized function $g(z)$. One could easily show that for any generic function $g(z)$ satisfying the condition $g(z=0) = 1$, the current value of $w_{\rm DE}$ will be 
\begin{eqnarray}
w_{\rm DE} (z=0) & = & -1 - \frac{1}{3 \ln (10)} ~\frac{d g(z)}{dz}\bigg|_{z=0}
\end{eqnarray}
which is below or above $-1$ depending on the sign of the derivative of the generic function at $z=0$. This actually opens the window of phenomenological dark energy models characterized by this general function.  

\begin{figure}[hbtp]
  \includegraphics[width=0.45\textwidth]{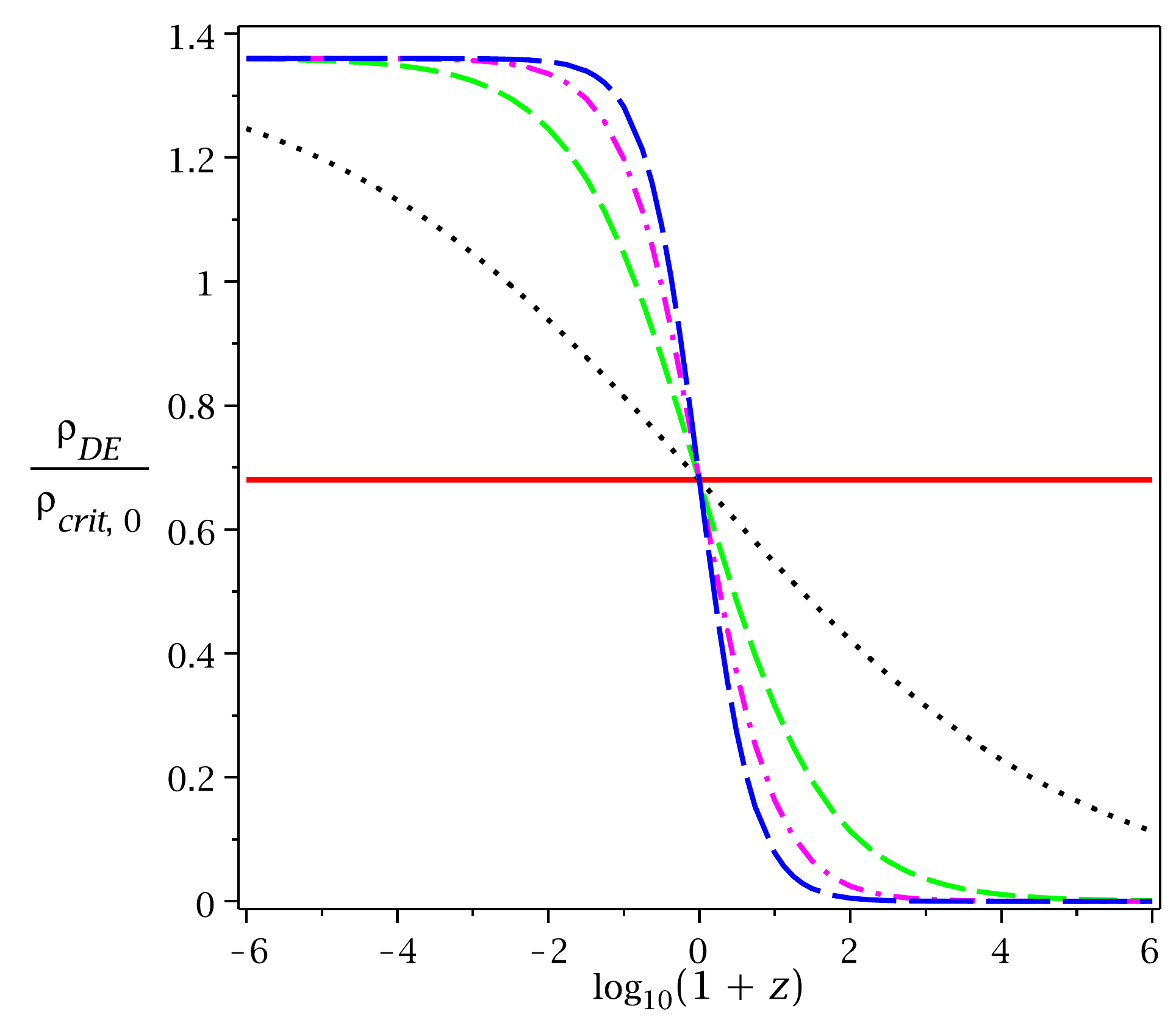} 
  \includegraphics[width=0.459\textwidth]{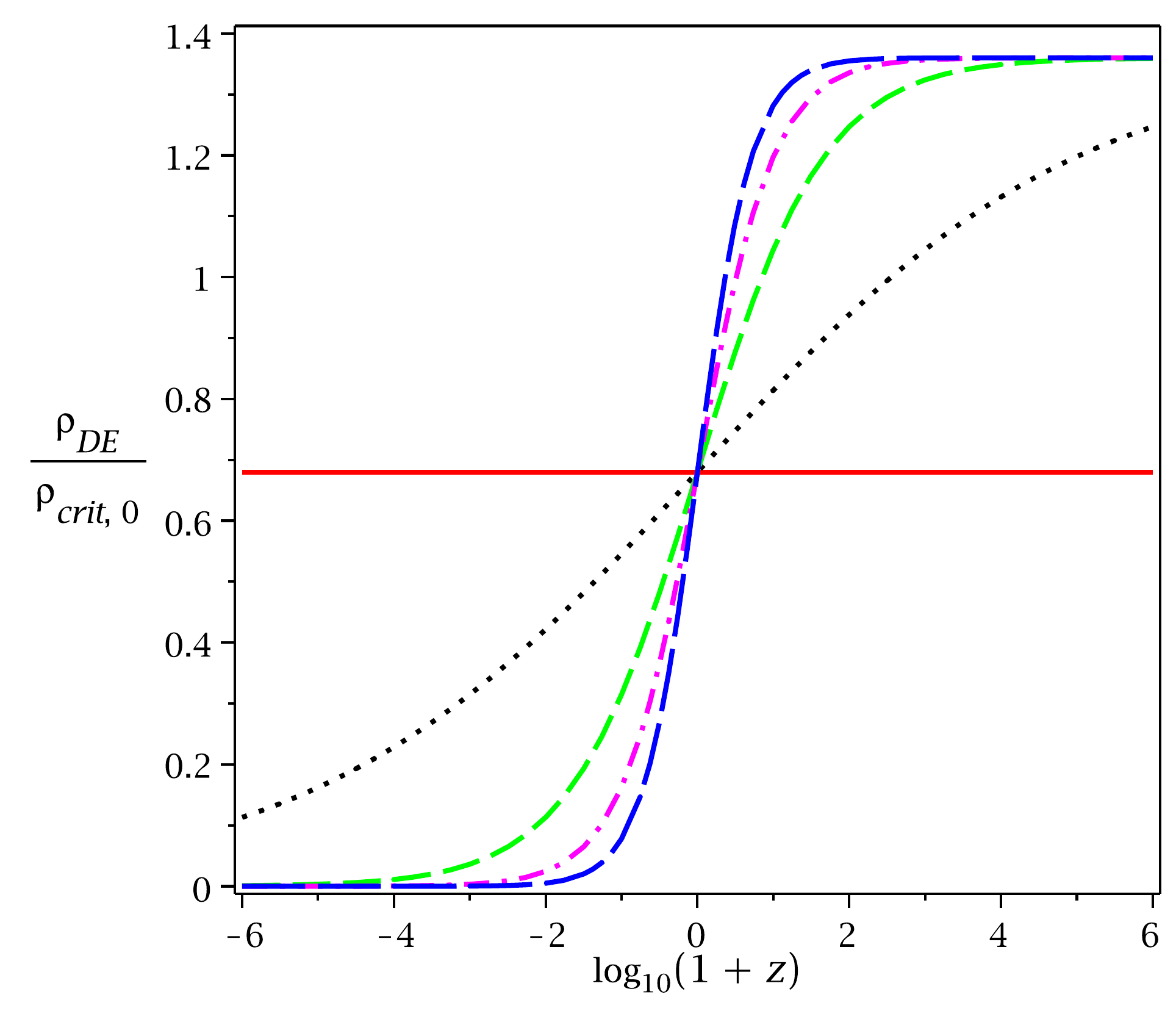} 
\caption{Evolution of the dark energy density with respect to the critical energy density (i.e., $\tilde{\Omega}_{\rm DE} (z) = \rho_{\rm DE}/\rho_{\rm crit,0}$) from early to future evolution of the universe, has been shown for different values 
of $\alpha$.  The left pane stands for $\alpha \geq 0$ and the right panel stands for $\alpha \leq 0$ where in both panels, $\alpha  =0$ represents the horizontal solid red line corresponding to the $\Lambda$CDM scenario with $\Omega_{\rm DE,0} = 0.68$. In the left panel, dot (black), dash (green), dash-dot (magenta) and long-dash (blue) respectively stands for $\alpha  = 0.2, \; 0.6,\; 1, 1.4$. Similarly, in the right panel, dot (black), dash (green), dash-dot (magenta) and long-dash (blue) respectively stands for $\alpha  = -0.2, \; -0.6,\; -1, -1.4$. }
\label{Fig:Omega}
\end{figure}

In Fig.~\ref{Fig:Omega}, we show the evolution of the dark energy density (with respect to the critical energy density)
for different values of $\alpha$. The left plot of Fig.~\ref{Fig:Omega} stands for $\alpha \geq 0$ and the right plot of Fig.~\ref{Fig:Omega} stands for $\alpha \leq 0$. The curve representing $\alpha  = 0$ in both the plots stands for $\Omega_{\rm DE,0} = 0.68$. We also note that in the left panel, the dot (black), dash (green), dash-dot (magenta) and long-dash (blue) curves respectively stands for $\alpha  = 0.2, \; 0.6,\; 1, 1.4$ while in the right panel, the same graphs, that means, the dot (black), dash (green), dash-dot (magenta) and long-dash (blue) curves respectively stands for $\alpha  = -0.2, \; -0.6,\; -1, -1.4$. From the left plot of Fig.~\ref{Fig:Omega} one can clearly see that the dark energy has an emerging nature, that means, in the past, dark energy has no effective presence but with the evolution of the universe, it emerges. The opposite scenario is visualized from the right plot of Fig.~\ref{Fig:Omega} (for $\alpha \leq 0$) which indicates that for the MEDE model, $\alpha > 0$ is a necessary condition for the emerging nature of dark energy.  Let us note that even if we change the value of $\alpha$, either in the positive or negative direction, the qualitative picture of the dark energy density remains same. Now, since the sign of $\alpha$ plays a crucial role in this context, specially its positive values allows us to have an emergent dark energy, however, in order to be more transparent, in the next section, we have considered both positive and negative values of  $\alpha$  for the statistical analysis and left this for the observational data to pick up the most acceptable value of $\alpha$.  This will determine the fate of the present dark energy model. 
 
\begin{figure*}
    \centering
    \includegraphics[width=0.45\textwidth]{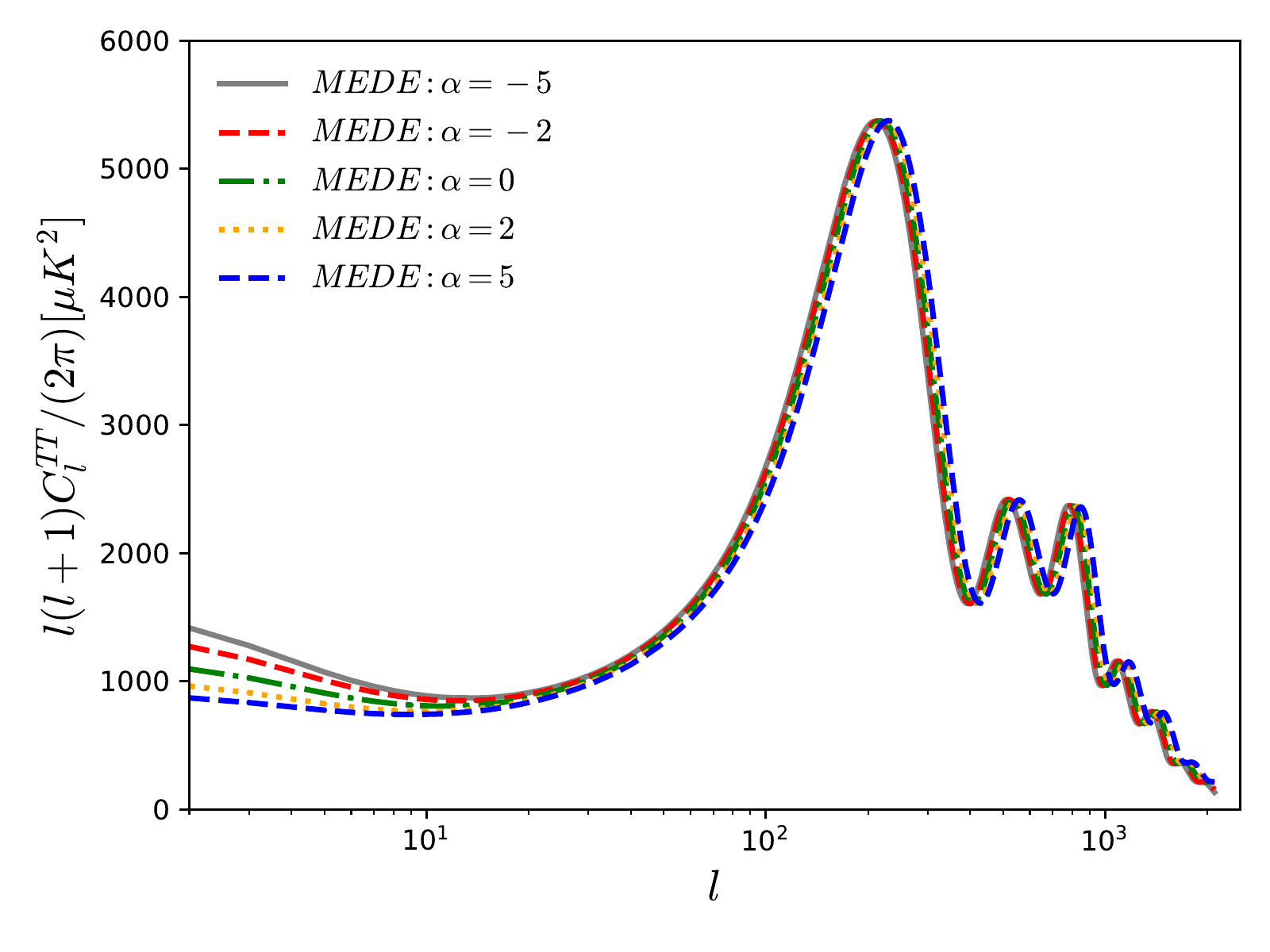}
    \includegraphics[width=0.45\textwidth]{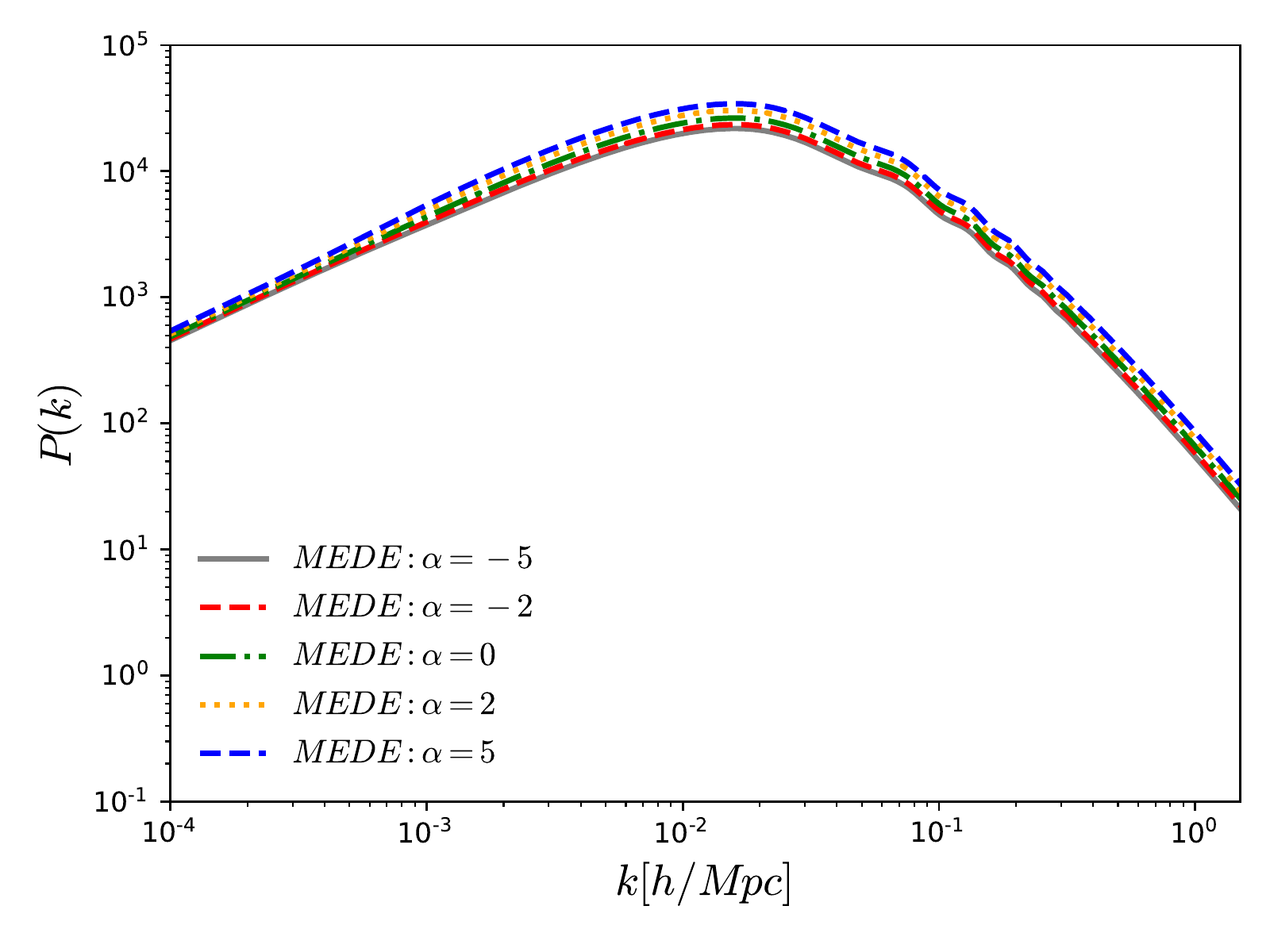}
    \caption{We show the CMB TT spectra (left plot) and matter power spectra (right plot) for the MEDE model using various values of the free parameter $\alpha$. }
    \label{fig:cmb+matter}
\end{figure*}

Let us now comment that an explicit relation between the pressure and energy density of the MEDE fluid can be established. In order to do so, let us write down the expression of the dark energy density as follows
\begin{eqnarray}
\rho_{\rm DE} & = & \rho_{\rm DE,0} \left(1 - \tanh \left[\alpha \log_{10} (1+z) \right] \right),
\end{eqnarray}
which can be obtained either by solving the conservation equation for dark energy or by simply looking at the definition of  $\tilde{\Omega}_{\rm DE}$. Here, 
$\rho_{\rm DE,0}$ is the present value of the DE density. 
Since the present fluid is barotropic, thus, using the relation $p_{\rm DE} = w_{\rm DE} \rho_{\rm DE}$, the pressure of the MEDE fluid can be obtained as 
\begin{eqnarray}
p_{\rm DE} & = & - \left(1 + \frac{2 \alpha}{3 \ln (10)} \right) ~\rho_{\rm DE} + 
\frac{\alpha}{3 \rho_{\rm DE, 0}\ln (10)} ~\rho_{\rm DE} ^2~.
\end{eqnarray}
So, the pressure of the MEDE fluid has a quadratic dependence on the energy density.  Let us close this section by investigating the behaviour of this model in the large scale of our universe. In order to do so, we investigate the CMB TT spectra and matter power spectra shown in the left and right plots of Fig. \ref{fig:cmb+matter}. The effects observed in the CMB TT and/or matter power spectra indicate the viability of the model because the instability in any model, if exists, is indicated through the blowing nature in one of those spectra. Here, we have analysed the CMB TT and matter power spectra considering both positive and negative values of $\alpha$. We find that the model does not indicate any instability in the large scale. Only one important observation comes through the behaviour of the model in the low multipole region, see the left plot of Fig. \ref{fig:cmb+matter}. We see that in the low multipole region (left plot of Fig. \ref{fig:cmb+matter}), $\alpha$ plays an important role while for the high multipole region, even if we increase the magnitude of $\alpha$, no notable feature is observed for a comment. 

\section{Observational data and Statistical Methodology}
\label{sec-observational data}

In this section we summarize the main observational datasets  used to constrain the MEDE model in this article and the statistical methodology. In what follows we describe the main observational datasets. 

\begin{enumerate}

\item \textbf{Cosmic Microwave Background (CMB)}: We use the CMB temperature and polarization angular power spectra {\it plikTTTEEE+lowl+lowE} for the latest Planck 2018 legacy release. 

\item \textbf{Baryon acoustic oscillations (BAO)}:  We use the same combination of BAO data considered by Planck 2018 in~\cite{Aghanim:2018eyx}, i.e.  6dFGS~\cite{Beutler:2011hx}, SDSS-MGS~\cite{Ross:2014qpa}, and 
BOSS DR12~\cite{Alam:2016hwk}.

\item \textbf{Hubble constant (R19)}: We consider a Gaussian prior $H_0 = 74.03 \pm 1.42$ km/s/Mpc at $68\%$ CL on the Hubble constant achieved by Riess et al. in \cite{Riess:2019cxk}, showing that there is a tension at $4.4 \sigma$ with the CMB value of $H_0$ assuming a minimal $\Lambda$CDM model.  

\item \textbf{Pantheon sample of Supernovae Type Ia}: We also employ the 1048 Supernoave Type Ia data points distributed in the redshift interval $z \in [0.01, 2.3]$, known as the Pantheon sample~\cite{Scolnic:2017caz}.  We note that Supernovae Type Ia data were the first astronomical data that signaled for an accelerating expansion.  

\end{enumerate}

We consider as a baseline a 7-dimensional parameter space described by the following parameters: the baryon energy density $\Omega_bh^2$ and cold dark matter energy density $\Omega_{c}h^2$, the ratio of the sound horizon at decoupling to the angular diameter distance to last scattering $100 \theta_{MC}$, the optical depth to reionization $\tau$, the spectral index and the amplitude of the primordial scalar perturbations $n_s$ and $A_s$, and the parameter $\alpha$ introduced in eq.~(\ref{alpha}). We assume flat uniform priors on all these parameters, as showed in Table~\ref{tab:priors}.

For the analyses, we considered a modified version the publicly available Markov Chain Monte Carlo code \texttt{CosmoMC}~\cite{Lewis:2002ah,Lewis:1999bs} package (see \url{http://cosmologist.info/cosmomc/}), in order to support the MEDE scenario. Moreover, this code is equipped with the last 2018 Planck likelihood~\cite{Aghanim:2019ame}, implementing an efficient sampling of the posterior distribution using the fast/slow parameter decorrelations \cite{Lewis:2013hha}. The code has, in addition, a convergence diagnostic based on the Gelman-Rubin statistics~\cite{Gelman:1992zz}. 

\begin{table}
\begin{center}
\begin{tabular}{|c|c|c}
\hline
Parameter                    & Prior\\
\hline 
$\Omega_{b} h^2$             & $[0.005,0.1]$\\
$\Omega_{c} h^2$             & $[0.001,0.99]$\\
$\tau$                       & $[0.01,0.8]$\\
$n_s$                        & $[0.8,1.2]$\\
$\log[10^{10}A_{s}]$         & $[1.6,3.9]$\\
$100\theta_{MC}$             & $[0.5,10]$\\ 
$\alpha$                     & $[-10,10]$\\
\hline
\end{tabular}
\end{center}
\caption{Flat priors assumed on the cosmological parameters associated with the MEDE model. Note that although from Fig. \ref{Fig:Omega}, $\alpha > 0$ is the required condition for the dark energy to be emerging in the late time and this is the focus of our work, however, from the statistical point of view, here we have allowed $\alpha$ to take both positive and negative values and left this for the observational data to pick up the best fit value of $\alpha$. }
\label{tab:priors}
\end{table}

\begingroup                                                 
\squeezetable                                                                                                                   
\begin{center}                                                                                                                  
\begin{table*}                                                                                                                   
\begin{tabular}{ccccccccccccccc}                                                                                                            
\hline\hline                                                                                                                    
Parameters & CMB+BAO & CMB+R19 & CMB+BAO+R19 & CMB+Pantheon &  CMB+BAO+Pantheon\\ \hline

$\Omega_c h^2$  & $    0.1197_{-    0.0013-    0.0025}^{+    0.0013+    0.0025}$ & $    0.1201_{-    0.0013-    0.0025}^{+    0.0013+    0.0025}$  & $    0.1206_{-    0.0013-    0.0026}^{+    0.0013+    0.0025}$ & $    0.1202_{-    0.0013-    0.0026}^{+    0.0013+    0.0026}$ &  $    0.1197_{-    0.0012-    0.0023}^{+    0.0012+    0.0023}$\\

$\Omega_b h^2$ & $    0.02239_{-    0.00015-    0.00028}^{+    0.00015+    0.00029}$  & $    0.02237_{-    0.00015-    0.00029}^{+    0.00014+    0.00030}$  & $    0.02235_{-    0.00014-    0.00027}^{+    0.00014+    0.00027}$ & $    0.02235_{-    0.00014-    0.00028}^{+    0.00014+    0.00028}$ &  $    0.02239_{-    0.00014-    0.00028}^{+    0.00014+    0.00029}$ \\

$100\theta_{MC}$  &  $    1.04095_{-    0.00031-    0.00060}^{+    0.00031+    0.00059}$  & $    1.04091_{-    0.00031-    0.00060}^{+    0.00031+    0.00063}$   & $    1.04087_{-    0.00030-    0.00059}^{+    0.00032+    0.00060}$ & $    1.04090_{-    0.00031-    0.00064}^{+    0.00034+    0.00059}$ &  $    1.04098_{-    0.00030-    0.00057}^{+    0.00029+    0.00059}$\\

$\tau$  & $    0.0549_{-    0.0079-    0.015}^{+    0.0077+    0.017}$   & $    0.0542_{-    0.0072-    0.014}^{+    0.0072+    0.015}$   & $    0.0531_{-    0.0081-    0.017}^{+    0.0084+    0.017}$ &  $    0.0540_{-    0.0076-    0.015}^{+    0.0076+    0.015}$ &  $    0.0553_{-    0.0083-    0.014}^{+    0.0073+    0.015}$ \\

$n_s$  & $    0.9659_{-    0.0041-    0.0084}^{+    0.0043+    0.0082}$   & $    0.9648_{-    0.0042-    0.0082}^{+    0.0042+    0.0083}$  & $    0.9637_{-    0.0043-    0.0080}^{+    0.0042+    0.0081}$&  $    0.9647_{-    0.0042-    0.0082}^{+    0.0042+    0.0082}$ &  $    0.9658_{-    0.0044-    0.0076}^{+    0.0040+    0.0080}$ \\

${\rm{ln}}(10^{10} A_s)$  & $    3.045_{-    0.016-    0.031}^{+    0.016+    0.033}$  & $    3.044_{-    0.015-    0.030}^{+    0.015+    0.031}$  & $    3.043_{-    0.016-    0.033}^{+    0.015+    0.032}$  & $    3.044_{-    0.015-    0.031}^{+    0.015+    0.031}$ &  $    3.046_{-    0.017-    0.030}^{+    0.015+    0.031}$ \\

$\alpha$  & $    0.17_{-    0.36-    0.83}^{+    0.44+    0.81}$   & $    1.33_{-    0.27-    0.53}^{+    0.27+    0.52}$  & $    0.88_{-    0.25-    0.56}^{+    0.28+    0.51}$ &  $    0.22_{-    0.25-    0.47}^{+    0.25+    0.47}$ &  $    0.18_{-    0.22-    0.45}^{+    0.24+    0.43}$ \\

$\Omega_{m0}$  & $    0.306_{-    0.014-    0.026}^{+    0.012+    0.027}$ & $    0.260_{-    0.011-    0.020}^{+    0.010+    0.020}$  & $    0.2818_{-    0.0094-    0.016}^{+    0.0077+    0.017}$  & $    0.307_{-    0.011-    0.020}^{+    0.010+    0.021}$ &  $    0.3059_{-    0.0078-    0.015}^{+    0.0075+    0.015}$ \\

$\sigma_8$ & $    0.819_{-    0.020-    0.040}^{+    0.020+    0.039}$  & $    0.875_{-    0.017-    0.033}^{+    0.017+    0.033}$  & $    0.852_{-    0.016-    0.033}^{+    0.016+    0.032}$ & $    0.822_{-    0.013-    0.025}^{+    0.014+    0.027}$ & $    0.819_{-    0.013-    0.027}^{+    0.013+    0.026}$ \\

$H_0 [{\rm km/s/Mpc]}$ & $   68.4_{-    1.5-    3.1}^{+    1.5+    3.0}$ & $   74.2_{-    1.4-    2.7}^{+    1.4+    2.7}$   & $   71.4_{-    1.0-    2.3}^{+    1.2+    2.0}$ & $   68.3_{-    1.0-    2.0}^{+    1.0+    2.1}$ &  $   68.33_{-    0.80-    1.6}^{+    0.85+    1.7}$ \\

$S_8$  & $    0.826_{-    0.013-    0.025}^{+    0.013+    0.026}$   & $    0.815_{-    0.014-    0.028}^{+    0.014+    0.029}$ & $    0.826_{-    0.013-    0.026}^{+    0.013+    0.027}$  & $    0.832_{-    0.015-    0.030}^{+    0.015+    0.029}$ &  $    0.827_{-    0.014-    0.026}^{+    0.014+    0.027}$ \\

\hline\hline                                                                                                                    
\end{tabular}                                                                                                                   
\caption{We show the 68\% and 95\% CL constraints on various free and derived parameters extracted from the Modified Emergent Dark Energy scenario for CMB+BAO, CMB+R19, CMB+BAO+R19, CMB+Pantheon, and CMB+BAO+Pantheon. }
\label{tab:set1}                                                                                                  
\end{table*}                                                                                                                     
\end{center}                                                                                                                    
\endgroup     
\begin{figure*}
\includegraphics[width=0.6\textwidth]{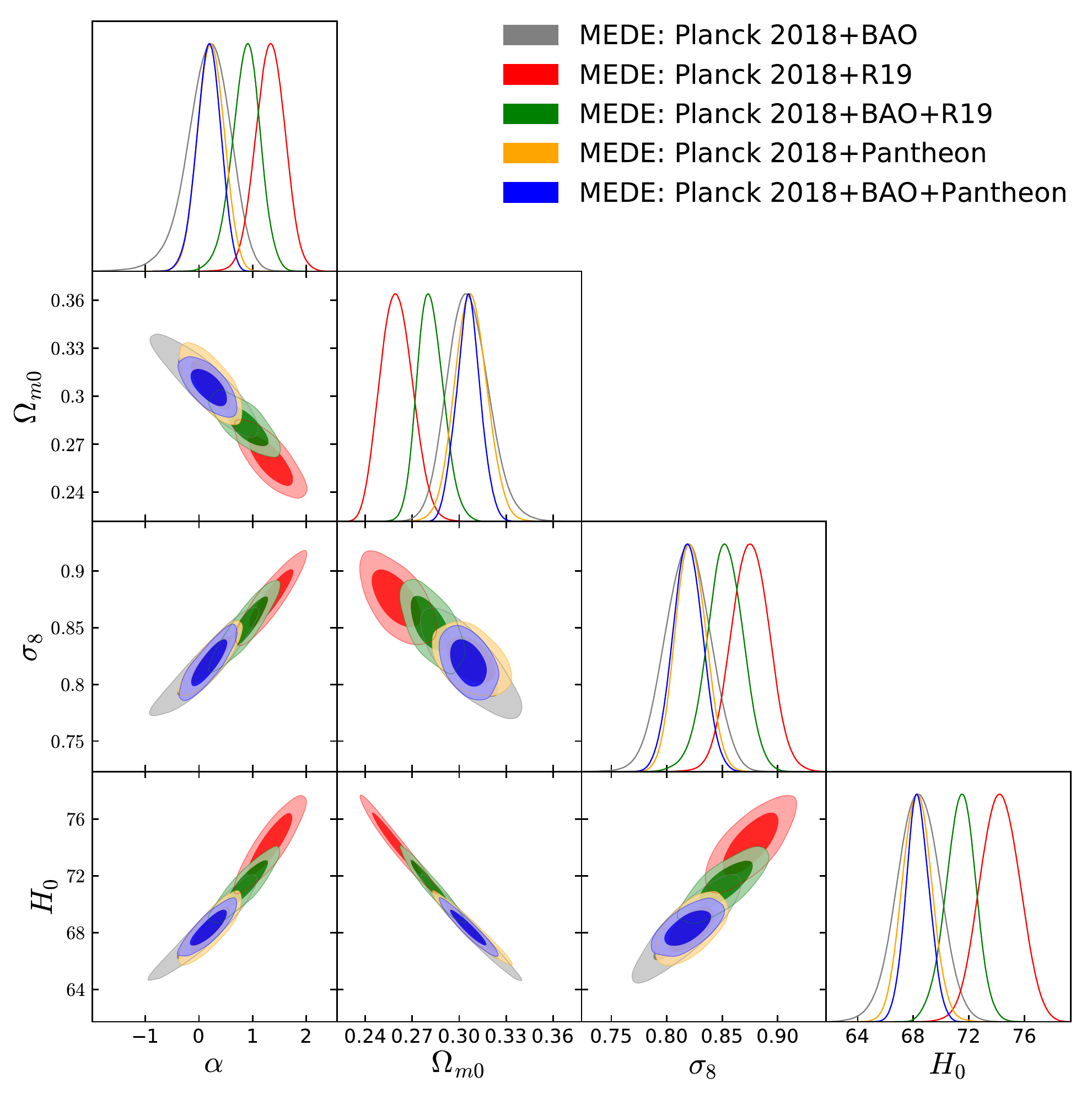}
\caption{1D marginalized posterior distributions and 2D $68\%$ and $95\%$~CL joint contours for several combinations of the parameters within the extended scenario MEDE using various cosmological datasets.}
\label{fig1}
\end{figure*}
\begin{figure*}
\includegraphics[width=0.6\textwidth]{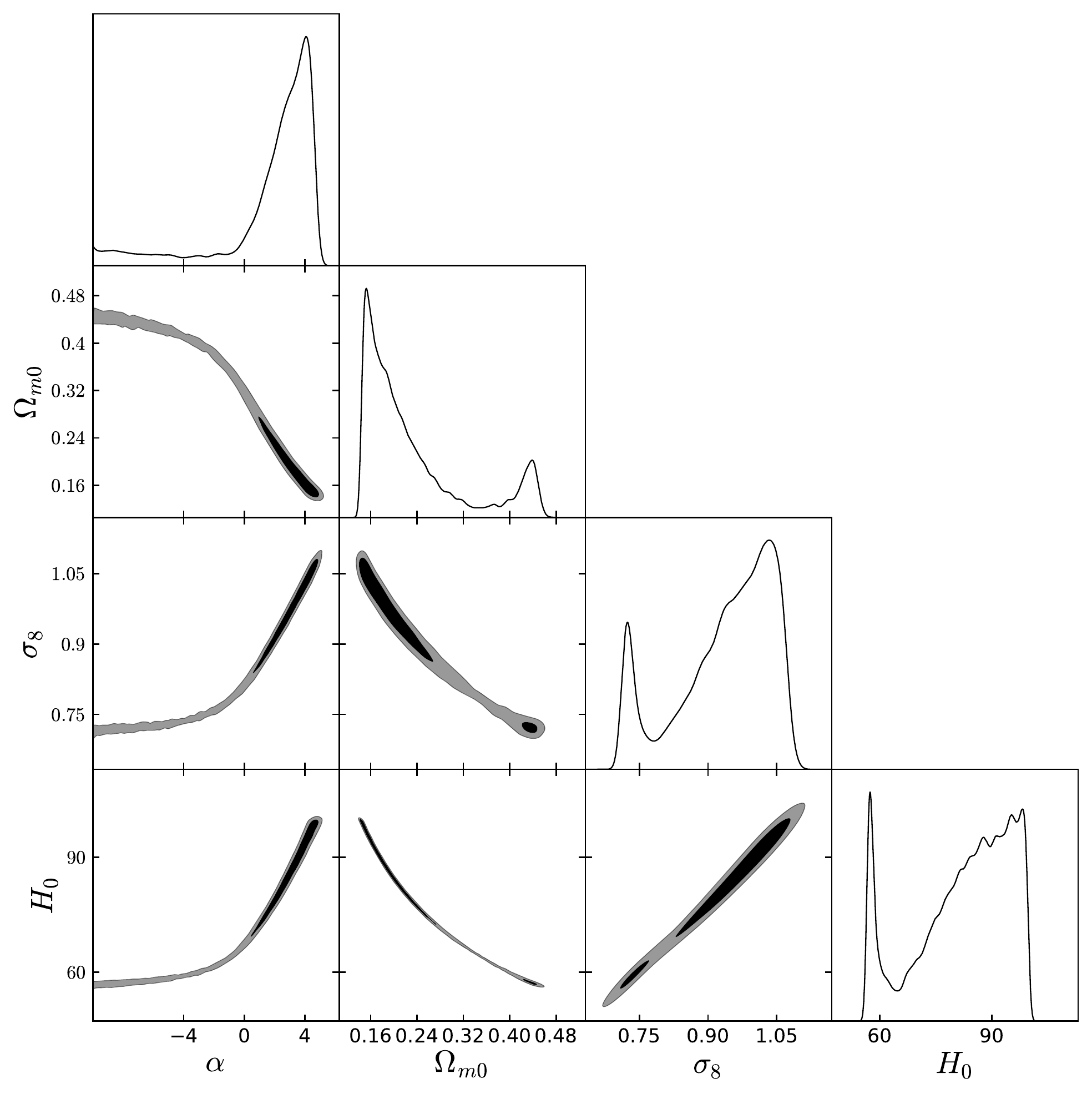}
\caption{1D marginalized posterior distributions and 2D $68\%$ and $95\%$~CL joint contours for several combinations of the parameters within the extended scenario MEDE for the CMB alone dataset showing the bimodal distribution. }
\label{fig2}
\end{figure*}

\section{Numerical Results and their implications}

\label{sec-results}

In this section we describe the fit of the data with the MEDE model and the cosmological constraints we have on the parameters.
In Table~\ref{tab:set1} we report the constraints at 68\% and 95\% CL on the 7 free parameter of the MEDE model and some derived quantities for several dataset combinations considered in this work. In Fig.~\ref{fig1} we show a triangular plot with the 1D posterior distributions and the 2D contour plots for some selected parameters.
We do not report the cosmological bounds for the CMB only case in the Table~\ref{tab:set1}, because, as we can see in Fig.~\ref{fig2}, this presents a bimodal distribution. Therefore, the CMB only dataset is not powerful enough in distinguishing one of the two peaks and the bounds are meaningless. 

We recall at this point that the MEDE scenario will be similar to a Dynamical Dark Energy model with only a free parameter $\alpha$ and dark energy equation of state given by eq.~(\ref{DEeos}). Moreover, with this parametrization we can recover the $\Lambda$CDM model when $\alpha = 0$, while we can re-obtain the PEDE case for $\alpha = 1$, a recently introduced model which has already been famous for its ability in solving the long standing Hubble constant tension~\cite{Li:2019yem,Pan:2019hac,Yang:2020myd}. Finally, we will have a phantom-like scenario if $\alpha>0$, and a quintessence-like DE if $\alpha<0$. Thus, the model has a cluster of properties quantified through the single parameter $\alpha$.

Since the 6 cosmological parameters in common with the $\Lambda$CDM model are qualitatively unaffected by the introduction of the MEDE scenario, here we will focus our discussion on the parameters $\alpha$ and $H_0$. This has the aim of understanding if the PEDE model is favoured with respect to the standard cosmological scenario, and if within this context, i.e. allowing for more freedom, we are still able to alleviate the $H_0$ tension.

With the CMB+BAO combination of data we find $\alpha=0.17^{+0.44}_{-0.36}$ at 68\% CL, in perfect agreement with zero and hence consistent with a cosmological constant scenario. However, the MEDE model allows $H_0=68.4\pm1.5$ km/s/Mpc at 68\% CL, larger than the original standard $\Lambda$CDM model, and able to alleviate the $H_0$ tension within 3 standard deviations. We can, therefore, combine CMB+BAO with the R19 prior on $H_0$, producing a shift of both $\alpha$ and the Hubble constant towards higher values, because of the positive correlation present between these two parameters (see Fig.~\ref{fig1}). We find, in fact, for CMB+BAO+R19 at 68\% CL $\alpha=0.88^{+0.28}_{-0.25}$  and $H_0=71.4^{+1.2}_{-1.0}$ km/s/Mpc, i.e. in agreement within 1$\sigma$ with the PEDE scenario \cite{Li:2019yem,Pan:2019hac,Yang:2020myd}.
The very same effect happens if BAO data are removed. Therefore, we have for CMB+R19 the Hubble constant $H_0$ fixed by the Gaussian prior R19, and a higher $\alpha=1.33\pm0.27$ at 68\% CL, recovering the PEDE model within 2$\sigma$, with a phantom-like preference for the DE.

Finally, we examined two other cosmological combinations, namely, CMB+Pantheon and CMB+BAO+Pantheon. 
For both the cases, we find that within 68\% CL, $\alpha  = 0$ is allowed, which means that the $\Lambda$CDM cosmology is in agreement within one standard deviation. However, concerning the estimations of the Hubble constant from both the datasets, $H_0 = 68.3 \pm 1.0$ km/s/Mpc (68\% CL, CMB+Pantheon) and $H_0 = 68.33_{-    0.80}^{+    0.85}$ km/s/Mpc (68\% CL, CMB+BAO+Pantheon), we find that they both are higher compared to the estimation by Planck 2018 alone within the minimal $\Lambda$CDM cosmology. This eventually infers that the tension on $H_0$ is slightly reduced, but still in disagreement with R19 above 3 standard deviations. In particular, for the CMB+Pantheon case, the $H_0$ tension is at 3.3$\sigma$ and for CMB+BAO+Pantheon, is instead at 3.4$\sigma$.  

In other words, we find that the PEDE model is favoured against the standard $\Lambda$CDM one, only where the R19 prior is involved, i.e. when is expected a larger $H_0$ value. In all the other dataset combinations, the standard $\Lambda$CDM scenario is still the favourite one in fitting the data.

\section{Summary and conclusions}
\label{sec-conclu}

Generalization of any theory or model is always welcome in science because any generalized version of a theory or model allows us to extend its barrier and hence new possibilities are expected. 
The theme of the present work is to investigate whether the recently introduced Phenomenologically Emergent Dark Energy (PEDE) model \cite{Li:2019yem} and the Generalized Emergent Dark Energy model~\cite{Li:2020ybr} can be modified in such a manner so that the modified version could be more interesting from the theoretical perspectives. Since the PEDE model is an excellent candidate to solve the $H_0$ problem \cite{Pan:2019hac,Yang:2020myd} quite nicely, hence, its generalization could certainly be interesting. Keeping all these in mind, we found that a modified version of the PEDE model which we call as  
{\it Modified Emergent Dark energy} (MEDE) model. The model recovers the PEDE model and the $\Lambda$CDM model as special cases quantified through an extra parameter $\alpha$.  Therefore,  unlike the PEDE model having exactly six parameters, here the MEDE model contains seven free parameters (six parameters from the PEDE model plus one parameter, $\alpha$ coming from its generalization) where for $\alpha  = 1$ we recover the PEDE model and for $\alpha  =0$ we recover the $\Lambda$CDM model.   The behaviour of the model in the large scale, as shown in Fig. \ref{fig:cmb+matter} is not abrupt for a wide range of the parameter $\alpha$. Only in the lower multipole region, $\alpha$ plays a significant role to distinguish between the curves while in the higher multipole region (see the left plot of Fig. \ref{fig:cmb+matter}), $\alpha$ does not seem to play any significant role.

In order to understand how the model behaves with the observational data, we constrained it using a series of cosmological observations including CMB from Planck 2018, BAO, R19 and Pantheon. In particular, we consider the following data combinations, CMB+BAO, CMB+R19, CMB+BAO+R19, CMB+Pantheon and CMB+BAO+Pantheon. 
The results are shown in Table \ref{tab:set1} and Fig. \ref{fig1}. We did not show the results for CMB alone because this dataset does not constrain this model well which is clearly visualized through Fig. \ref{fig2} showing the bimodal distribution. Our analyses are very transparent and robust. 
We find that unlike the PEDE scenario where the $H_0$ problem was solved quite satisfactorily, here within this MEDE scenario, the tension on $H_0$ is not solved, but it is certainly slightly alleviated. However, this varies from one combined dataset to another. Therefore, the favoured scenario in fitting the data is always the $\Lambda$CDM model, with the exception of the cases in which the R19 prior, and a larger $H_0$ value, is involved.

In summary, the present work offers a novel approach to construct various PEDE models using a simple condition on the generic function $g (z)$ and opens a new window of a cluster of new cosmological models. We hope that the investigators like us will be motivated to explore this sector.

\section*{ACKNOWLEDGMENTS}
HBB gratefully acknowledges the financial support from University of Sharjah (grant number V.C.R.G./R.438/2020). WY acknowledges the support by the 
National Natural Science Foundation of China under Grants 
No. 11705079 and No. 11647153. SP was supported by the Mathematical Research Impact-Centric Support Scheme 
(MATRICS), File No. MTR/2018/000940, given by the Science and Engineering Research Board 
(SERB), Govt. of India.
EDV was supported from the European Research Council in the form of a Consolidator Grant 
with number 681431.


\bibliography{biblio}

\end{document}